\documentstyle[epsfig]{aipproc}

\begin{document}
\title{Performance of the STACEE Atmospheric Cherenkov Telescope}

\author{D.A.~Williams,$^1$ D.~Bhattacharya,$^2$ L.M.~Boone,$^1$ 
M.C.~Chantell,$^3$ Z.~Conner,$^{3*}$ C.E.~Covault,$^3$ 
M.~Dragovan,$^{3\dagger}$ P.~Fortin,$^4$ D.~Gingrich,$^5$ 
D.T.~Gregorich,$^6$ D.S.~Hanna,$^4$ G.~Mohanty,$^2$ R.~Mukherjee,$^7$ 
R.A.~Ong,$^3$ S.~Oser,$^{3\ddagger}$ K.~Ragan,$^4$ R.A.~Scalzo,$^3$
D.R.~Schuette,$^3$\\
C.G.~Th\' eoret,$^4$ T.O.~T\" umer,$^2$ F.~Vincent,$^4$
J.A.~Zweerink$^2$}
\address{
$^1$Santa Cruz Institute for Particle Physics,
Univ. of California, Santa Cruz, CA 95064, USA
\\
$^2$Institute of Geophysics and Planetary Physics, Univ.\ 
of California, Riverside, CA 92521, USA\\
$^3$Enrico Fermi Institute, Univ.\ of Chicago, Chicago, IL 60637, USA\\
$^4$Dept.\ of Physics, McGill Univ., Montreal, 
Quebec H3A 2T8, Canada\\
$^5$Centre for Subatomic Research, Univ.\ of Alberta, Edmonton, 
Alberta T6G 2N5, Canada\\
$^6$Dept.\ of Physics \& Astronomy, California State Univ.,
Los Angeles, CA  90032, USA\\
$^7$Dept.\ of Physics \& Astronomy, Barnard Coll. \& Columbia
Univ., New York, NY 10027, USA\\
$^*$Present address: Dept.\ of Physics, George Washington Univ., Washington, D.C. 20077, USA\\
$^\dagger$Present address: Jet Propulsion Laboratory, Pasadena, CA 91109, USA\\
$^\ddagger$Present address: Dept.\ of Physics \& Astronomy, 
Univ.\ of Pennsylvania, Philadelphia, PA 19104, USA}

\maketitle

\begin{abstract}
The Solar Tower Atmospheric Cherenkov Effect Experiment (STACEE)
is 
located at the National Solar Thermal Test Facility of 
Sandia National Laboratories in Albuquerque, New Mexico, USA.
The field of solar tracking mirrors (heliostats)
around a central receiver tower is
used to direct Cherenkov light from atmospheric showers onto secondary
mirrors on the tower, which in turn image the light onto cameras of
photomultiplier tubes.
The STACEE Collaboration has previously reported a detection of the
Crab Nebula with approximately 7 standard deviation significance, 
using 32 heliostats (STACEE-32).  This result
demonstrates both the viability of the technique and the suitability
of the site.  We are
in the process of completing an upgrade to 48 heliostats (STACEE-48) 
en route to an eventual configuration using 64 heliostats (STACEE-64)
in early 2001.  In this paper, we summarize the results obtained
on the sensitivity of STACEE-32 and our expectations for STACEE-48
and STACEE-64.
\end{abstract}

Astrophysical measurements in the 10 to 300 GeV region of the gamma-ray 
spectrum have 
proved elusive to both ground-based and satellite techniques.
Satellite detectors have been too small to detect the faint high-energy
fluxes.  Ground-based detectors can achieve large collection areas, but 
rely on detecting the secondary products of 
extensive air showers in the atmosphere.
The imaging atmospheric Cherenkov technique has been quite successful
above a few hundred GeV\cite{ong}, 
the lowest threshold at which the Cherenkov light
image stands out above the night-sky background.  
Builders of the next generation of imaging telescopes aspire
to thresholds as low
as 10 GeV\cite{nextgeneration}.  
STACEE is a wavefront-sampling Cherenkov telescope 
which, by using an existing solar facility, has a large mirror area to
achieve a low threshold and
has been implemented in a comparatively short time.

STACEE is situated at the National Solar Thermal Test Facility (34.962$^\circ$
N, 106.509$^\circ$ W, 1705 m above sea level) near Albuquerque, New Mexico.
The facility has 212 heliostat mirrors, each about 37 m$^2$ area, designed
to track the Sun and direct the image onto a central tower.  As part of the
STACEE design, we performed a detailed evaluation of the suitability of the
site and the infrastructure.  The results have been summarized 
previously \cite{staceenim}.  
For STACEE, the heliostats are used at night to 
direct Cherenkov light from gamma-ray showers in the atmosphere onto
secondary mirrors on the central tower.  The secondaries focus the light
onto cameras of photomultiplier tubes.  The secondary mirrors, phototube
cameras, trigger and read-out electronics, and data acquisition system 
must be provided
by the STACEE group, as well as software for retargeting the heliostats.
The heliostats and their control system are maintained by the staff of the
Test Facility.

We are working towards instrumenting 64 of the heliostats for use in 
STACEE.  Since the time of flight for light from each heliostat to the
secondary on the tower is different, the light from each heliostat must
be focussed on a separate phototube in the camera in order to preserve
the timing information.  Once we had completed the instrumentation of
32 heliostats (``STACEE--32''), 
we took a substantial data set on the Crab nebula and
pulsar.  We presently have 48 heliostats instrumented (``STACEE-48''),
and expect to complete the preparation of the final 16 to complete
STACEE--64 in early 2001.
The results from the analysis of the STACEE--32 Crab data\cite{oser,staceeapj} 
provide the most 
comprehensive information we have so far on STACEE performance.  We
will first summarize those results.  We refer the reader to the references
for the full details.  We will then discuss
our expectations for STACEE--48 and STACEE--64.

STACEE--32 was used to collect data from the Crab from November 1998 to
February 1999.  In addition to any possible gamma-ray signal from the
Crab, there is a large isotropic background of showers from charged 
cosmic rays, especially protons.  To account for this background, we
compare the event rate tracking the direction of the Crab (``on-source'')
to the rate tracking the same path on the sky in local coordinates,
7.5\arcdeg\ to the east or west of the Crab position in right ascension
(``off-source'').  Data were collected in one hour cycles, consisting of
two 28 minute runs (one on-source and one off-source) with 2 minutes 
between runs for slewing to the new position.  Weather conditions were
closely monitored, and only on/off pairs during good, stable conditions 
were used in the analysis.  In some cases a questionable period was 
identified in one run of a pair.  That period and the corresponding period
in the second run of the pair were both removed, in order to have
matched exposure in the on and off data sets.
Events collected during periods satisfying the selection criteria for 
good conditions show a 5.3 standard deviation excess in the on-source
events.  

\begin{table}
\begin{center}
\begin{tabular}{rcccc}
\tableline
\tableline
Quantity   &    Nov 98  &  Dec 98 & Jan/Feb 99 & Total \\
\tableline
On-Source Time (s)   & 56056 & 51239 & 48040 & 155335 \\
On-Source Events            & 76235 & 55634 & 51046 & 182915 \\
Off-Source Events           & 74686 & 54342 & 49825 & 178853 \\
Significance     & $4.0\sigma$ & $3.9\sigma$ & $3.8\sigma$ & $6.8\sigma$\\Excess 
Rate (min$^{-1}$) & $1.7 \pm 0.4$ & $1.5 \pm 0.4$ &
$1.5 \pm 0.4$ & $1.57 \pm 0.23$ \\
\tableline
\end{tabular}
\caption{Event excesses from the direction of the Crab.
The January and February data have
been combined, because fewer runs were taken in those months.\label{crabtable}}
\end{center}
\end{table}

All of these events were then 
reconstructed using the information on the time
of arrival of the Cherenkov light pulse at each heliostat.  Some events
that result from night-sky background fluctuations can be eliminated
by reimposing the trigger condition in software.  The 32 heliostats are 
organized into clusters of 8.  In order to trigger, there must be at least
3 of the 4 clusters which have 5 or more hits each.  This condition is 
implemented in hardware using programmable delays and coincidence modules.
It can be replicated somewhat more precisely in software using the recorded
hit times.

The Cherenkov wavefront for a $\approx$100 GeV gamma-ray shower can 
be approximated as a sphere centered at the shower maximum.  The wavefront
for proton-initiated showers is much more irregular.  We use this as a
means of preferentially selecting gamma-ray showers and suppressing the
background.  We fit the timing information for each shower to a sphere,
and keep only those events for which the $\chi^2$ per degree of freedom
is less than $1.0$.  The results for the number of events passing these 
requirements are summarized in Table~\ref{crabtable}.  
The excess has been enhanced to 6.8 standard deviations by the
imposition of the timing requirements, an effect which would be
expected if the excess is the result of a gamma-ray signal.
The excess rate is steady throughout the data set, which supports
the premise of a steady signal.

\begin{figure*}[t]
\centerline{
\epsfig{file=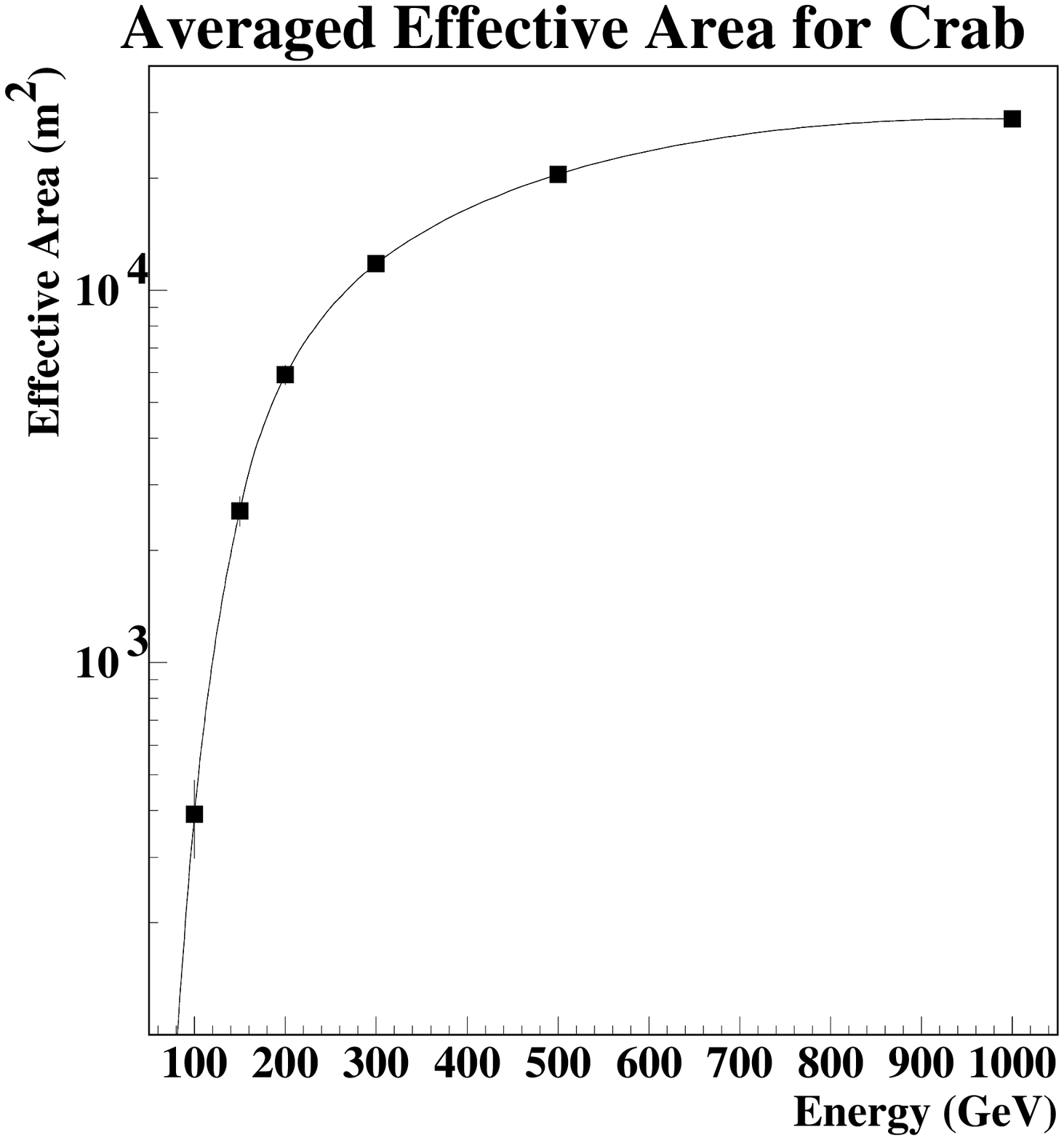,height=2.65in,bbllx=-150pt,bblly=0pt,bburx
=725pt,bbury=595pt,clip=.}\hskip0.5in
\epsfig{file=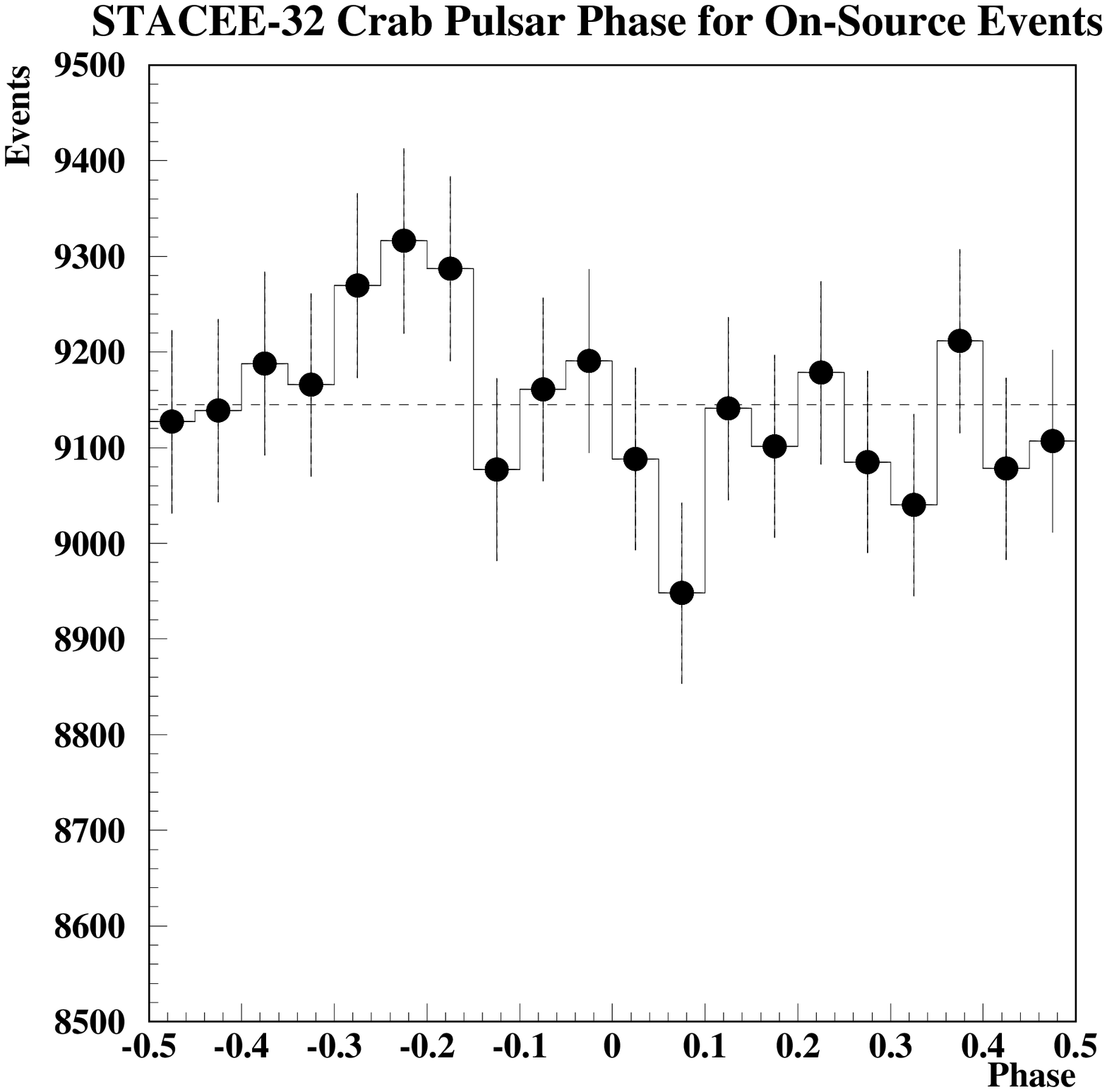,height=2.65in,bbllx=100pt,bblly=0pt,bburx
=725pt,bbury=595pt,clip=.}}
\smallskip
\caption{On the left, the Effective Area for STACEE--32.  On the right, the
STACEE--32 Crab Pulsar Phase Histogram.
(\copyright 2001. The American Astronomical Society. \protect\cite{staceeapj})}
\label{figure}
\end{figure*}

In order to determine the flux corresponding to the observed excess, we have
done a thorough study of the STACEE--32 performance.  We have used a CCD 
camera to record images of the Sun on the front face of the solar tower
and images of the full Moon in the focal plane of the PMT camera.  These
images have been used to evaluate the collection efficiency of the optics.
We have made a detailed simulation of the electronics and trigger, including
the measured shape and variety of single photoelectron pulses and the effects
of night-sky background light.  We use the program MOCCA to do a full 
simulation of the shower development and Cherenkov light 
production in the atmosphere.
The effective area that we obtain for STACEE--32, averaged over our 
Crab exposure, is shown in Fig.~1.  Multiplied by a differential
source spectrum proportional to $E^{-2.4}$, we obtain a spectral energy
threshold, $E_{th}$ 
(defined as the peak of the observed differential spectrum) of
190$\pm$60 GeV.  We measure the integral flux of gamma rays 
from the Crab Nebula
above $E_{th}$ to be
\begin{equation}
I (E > E_{th}) = (2.2 \pm 0.6 \pm 0.2)~
\times~10^{-10}\ {\rm photons~cm^{-2}~s^{-1}}.
\end{equation}
The first error is statistical, and the second error is the systematic
error on the flux itself, not including the effects of uncertainty in the
energy threshold.  The uncertainty in $E_{th}$ does not change the flux
value itself, but reflects the uncertainty in the energy above
which the flux is integrated.

We have searched for evidence of pulsed emission in our detected excess
from the Crab.  The phase distribution of on-source events is shown in 
Fig.~1, where 0.0 corresponds to the phase of the main radio pulse.  The
distribution is consistent with being flat, and we place an upper limit
on the fraction of the excess in the phase intervals where EGRET \cite{nolan} 
has observed pulsed emission at lower energy. 
Those intervals are 0.94--0.04 and 0.32--0.43.  The STACEE--32 limit
on the pulsed fraction in these intervals is $<$5.5\%\ at the 90\%\ 
confidence level.

Based on our work with STACEE--32, we can begin to extrapolate to the 
performance of STACEE--48 and STACEE--64.  The most obvious upgrade is
the addition of more heliostats, improving the light collection and 
the ability to pick out faint flashes.  We are making several other
improvements ``behind the scenes'' as well:
\begin{itemize}
\item
We are adding 1 GHz Flash ADC's on all channels for better time and
pulse amplitude resolution, especially for small pulses.
\item
We have a new trigger with a tighter coincidence window (presently 8 ns,
but capable of being reprogrammed even tighter) with lower dead time at
high rates.
\item
We have improved the focusing and alignment of the heliostats to increase
the amount of light reaching the secondary mirrors.
\item
We have applied a black sealant to the ground underneath the heliostats to
reduce the albedo.
\end{itemize}
We are still in the process of evaluating the improvement in sensitivity
from all of these
changes.  We expect to achieve a spectral energy threshold below 100 GeV
for STACEE--48, and near 50 GeV for STACEE--64, with much increased effective
area compared to STACEE--32 in both cases.   

This work was supported in part by the National Science Foundation, the
Natural Sciences and Engineering Research Council, Fonds pour la Formation
de Chercheurs et l'Aide \` a la Recherche, the Research Corporation, and 
the California Space Institute.

\end{document}